\documentstyle[12pt]{article}
\def\boxit#1{\kern4pt\vbox{\hrule\hbox{\vrule\kern8pt
\vbox{\kern8pt#1\kern8pt}\ern8pt\vrule}\hrule}}

\def\and{ {\rm and} }

\begin{document}
\onecolumn

\vfill

\begin{center}
{\Large \bf 
Test of Special Relativity and Equivalence principle from 
K Physics} \\ \vfill 
T. Hambye$^{(a)}$\footnotemark\footnotetext{email: 
hambye@hal1.physik.uni-dortmund.de}, 
R.B. Mann$^{(b)}$\footnotemark\footnotetext{email: 
mann@avatar.uwaterloo.ca}
and U. Sarkar$^{(c)}$\footnotemark\footnotetext{email: utpal@prl.ernet.in}\\
\vspace{2cm}
(a) Institut f\"{u}r Physik, Universit\"{a}t Dortmund,
D-44221 Dortmund, Germany \\
(b) Physics Department, University of Waterloo,
Waterloo, Ontario, Canada N2L 3G1\\
(c)Theory Group, Physical Research Laboratory,
Ahmedabad, 380 009, India\\
\end{center}

\vfill

\begin{abstract} 

A violation of Local Lorentz Invariance (VLI) and hence the 
special theory of relativity or a violation of equivalence 
principle (VEP) in the Kaon system can, in principle, induce 
oscillations between $K^0$ and $\bar{K}^0$. We construct a 
general formulation in which simultaneous pairwise 
diagonalization of mass, momemtum, weak or gravitational 
eigenstates is not assumed.
We discuss this problem in a general way and point out that, as 
expected, the VEP and VLI contributions are indistinguishable. 
We then insist on the fact that VEP or VLI can occur even when CPT is 
conserved. A possible $CP$ violation of the 
superweak type induced by VEP or VLI is introduced and discussed. We 
show that the general VEP mechanism (or the VLI mechanism, 
but not both simultaneously), with or without conserved CPT,
could be clearly tested experimentally through 
the energy dependence of the $K_L-K_S$ mass difference and of 
$\eta_{+-}$, $\eta_{00}$, $\delta$.  Constraints imposed by 
present experiments are calculated. 

\end{abstract}

\vfill

\section{Introduction.}

A few of the basic building blocks of particle physics are the 
assumptions that nature preserves local Lorentz invariance and 
hence the special theory of relativity, the product of the discrete 
symmetries CPT and the equivalence principle. It is also true 
that to date we have not seen any violation of any of these 
laws. In recent times many new attempts have been made to obtain
new and quantifiable information on the degree of validity of these 
basic laws. It is in this connection that we plan to 
investigate the Kaon--system. 

Many experiments have tested the special theory of relativity to a 
high degree of precision \cite{spexp}. These experiments probe 
for any dependence of the (non-gravitational) laws of physics 
on a laboratory's  position, orientation or velocity relative 
to some preferred frame of  reference, such as the frame in 
which the cosmic microwave background is isotropic. Failure to observe
such dependence further enhances the validity of 
(respectively) Local Position Invariance and Local Lorentz 
Invariance (LLI), and hence of the Einstein Equivalence Principle (EEP)
\cite{will}. However, these empirical results have been obtained primarily
in the baryon-photon sector of the standard model. There 
is no logically necessary reason to conclude from these results
that the special theory of 
relativity must be valid in all sectors of the standard model 
of elementary particle physics.  Its validity must be 
empirically checked for each sector (gauge boson, neutrino,
massive lepton, {\it etc.}) separately \cite{cathugh}.

A characteristic feature of LLI violation (VLI) is that every 
species of matter has its own maximum attainable speed. This 
yields several novel effects in various sectors of the standard 
model \cite{cathugh}, including vacuum Cerenkov radiation 
\cite{gasp}, photon decay \cite{cole} and neutrino oscillations 
\cite{glash,utrobb,jonas}. Recently we extended these arguments and 
pointed out that violation of special relativity will in 
general induce an energy dependent $K_L - K_S$ mass difference 
\cite{hambye}; an empirical search for such effects can 
therefore be used to obtain bounds on VLI in the Kaon 
sector of the standard model. As we shall discuss later VLI 
in the kaon sector can occur in a manner that may or may not 
violate CPT.

The  EEP implies   universality  
of gravitational  coupling  for all  forms of  mass-energy,  
thereby ensuring  that  spacetime is  described  by a unique  
operational geometry.  An extreme converse of this principle is
that every form of stress-energy couples to its own metric, so
that the Lagrangian for the standard model is modified to be
one of the form
\begin{equation}\label{introeq}
{\cal L} = {\cal L}_G(g^I) + \sum_I{\cal L}_M(g^I,\Phi_I) + {\cal L}_C
\end{equation}
where each matter field $\Phi_I$ couples to its own metric $g^I_{\mu\nu}$.
The gravitational Lagrangian density ${\cal L}_G$ describes the behaviour 
of all of these metrics in the absence of any matter fields. The Lagrangian
density ${\cal L}_C$ describes the interaction between the different 
matter fields; it will in general include at least some subset of the 
metric fields $g^I_{\mu\nu}$.   Although such a Lagrangian is generally 
covariant, spacetime no longer has a unique operational
geometry, since clocks and measuring rods constructed out of different
types of matter fields will in general yield different results for a
given set of experiments that depend on the choice of coordinate frame.
Furthermore, while it is possible for any given metric $g^I_{\mu\nu}$ 
to interpret  a diffeomorphism of the manifold as a gauge transformation of the 
linearized tensor $h^I_{\mu\nu} = g^I_{\mu\nu} - \bar{g}^I_{\mu\nu}$
where $\bar{g}^I_{\mu\nu}$ is some reference metric (typically chosen
to be a flat metric),  this cannot be done simultaneously for
all the metrics (unless they are all the same).  This means that the
spin modes of all the other metrics will in general be excited.  It is
then a theoretical challenge to ensure that the excitations of the
additional degrees of freedom of the other metric do not yield 
unacceptable pathologies such as runaway negative energy solutions,
tachyons, etc.  One might imagine doing this by giving, say, 
the gravitons associated with the metrics a tiny mass, save for the
metric associated with ordinary stable matter.  More general
theoretical mechanisms than that given in (\ref{introeq})
can also be considered: for example some of the metrics may not
be describable by second rank tensor fields, or some sectors of
the theory may not even be Lagrangian-based.  For an overview
and further discussion of the different possibilities, 
see ref. \cite{will}.

From an empirical perspective, the validity of the EEP
must therefore be checked sector-by-sector in the standard
model, since it cannot be imposed on grounds of logic.
Although the EEP has been tested to  
impressive  levels of precision, virtually all such tests have 
been carried out with matter  fields.  The  possibility  that 
matter and antimatter may have different gravitational 
couplings remains a fascinating open question.  The strongest 
bound on matter-antimatter gravitational universality  comes  
from  the  $K^0 - \bar{K}^0$ system.  Recent studies of this 
system have considered a straightforward violation of the weak 
equivalence principle (WEP) in which it is assumed that $K^0 - 
\bar{K}^0$ mass and gravitational eigenstates can be 
simultaneously diagonalised but with differing eigenvalues 
({\it i.e.} differing $K^0$ and $\bar{K}^0$ masses) 
\cite{good,Kenyon,hughes}, in which case violation of  
gravitational universality also means violation of CPT.

However, more generally, a violation of the EEP (VEP) in the 
Kaon system will not assume simultaneous pairwise 
diagonalization of mass, gravitational or weak eigenstates. We 
shall consider in this article the consequences of such a 
general VEP mechanism, showing that it can provide a source 
of $CP$ violation whilst conserving CPT. In this context, 
previously investigated mechanisms of EEP violation in the Kaon 
system may be considered either as special cases of maximal 
$CPT$ violation in the gravitational sector \cite{good,Kenyon,hughes} 
or else CPT conserved VEP, which is the other extreme case 
\cite{nacht}. Our analysis is more general, including 
all the earlier analyses as special cases and in addition allows us 
to compare with the VLI bounds. We consider constraints imposed 
on this general VEP mechanism by present experiments.

In section 2 and 3 we derive the general mass matrix including VLI and VEP 
effects respectively. 
In both sections we  
point out that VLI (VEP) allows for a phase $\alpha_v$ ($\alpha_G$) 
responsible for $CP$ violation whilst conserving $CPT$.
At the end of section 3 we compare both general mass matrices
noticing that VEP and VLI effects are 
indistinguishable as expected
\cite{will} (for the neutrino sector this similarity was 
pointed out in ref.\cite{glash}). In section 4 we discuss the general 
case where these phases are taken to be 0.  We consider first the
$CPT$-conserving case  and examine the energy dependence VLI and VEP 
induce in the mass difference $m_L - m_S$. Constraints on VEP parameters 
(and hence on VLI parameters) from experiments on $m_L - m_S$ 
are discussed. 
We also give constraints on the interesting maximally $CPT-$violating case
where matter or antimatter states are the velocity or gravitational
eigenstates with differing eigenvalues.
Then in section 5 we discuss the effect of the 
phases $\alpha_v$ and $\alpha_G$ and constraints on VEP and VLI
parameters from 
$CP$ violation experiments. We point out that the $CP$ violation 
induced by VEP or VLI is of the superweak type and has an inherent 
energy dependence. Consequently, although this mechanism cannot 
fully account for observed $CP$ violation in the Kaon system, 
it yields a definite testable prediction for the energy 
dependence of $CP$ violation parameters. This can then be used 
to put a qualitatively new bound on VEP or VLI.
We summarise our results in section 6.

\section{Violation of LLI.}

The maximum attainable velocities of particles and antiparticles 
can differ if there is violation of LLI \cite{cole}. Here we 
take a phenomenological 
approach to this problem and assume that neither the mass nor the weak 
eigenstates are a priori simultaneously diagonalisable with the 
momemtum eigenstates. 

Then the 
general form of the effective Hamiltonian associated with the 
Lagrangian in the $(K^0 \hskip .10in \bar{K}^0)$ basis will be 
\begin{equation}
H = U_W H_{SEW} U_W^{-1} + U_v H_v U_v^{-1} \label{h}
\end{equation}
with, 
\begin{equation}
H_{SEW} = \frac{(M_{SEW})^2}{2 p} = \frac{1}{2 p} {\pmatrix{ 
m_1 - i \frac{\Gamma_1}{2} & 0 \cr 0 & m_2 -i \frac{\Gamma_2}
{2}}}^2 \label{hsew}
\end{equation}
and
\begin{equation}
H_v = \pmatrix{ v_1 & 0 \cr 0 & v_2} p \label{hg}
\end{equation}
to leading order in $\bar{m}^2/p^2$ with p the momentum and 
$\bar{m} = (m_1 + m_2)/2$ the average mass. From now on we
define $\delta X \equiv(X_1-X_2)$, $\bar{X} \equiv (X_1+X_2)/2$
for any quantity $X$.  
$H_{SEW}$ refers to the
strong and electroweak part of the hamiltonian. 
The constants $v_1$ 
and $v_2$ correspond to the maximum attainable speeds of each 
eigenstate. If special relativity is valid within the Kaon 
sector these are both equal to their average $\bar{v}=(v_1 + 
v_2)/2$, which we normalize to unity.  Hence 
$v_1 - v_2 = \delta v$ is a measure of VLI in the 
Kaon sector.  If $\bar{v}$ corresponds to the speed of 
electromagnetic radiation then special relativity is valid 
within the Kaon--photon sector of the standard model. 
In the limit $v_1=v_2$, 
$m_{1,2}$  and  $\Gamma_{1,2}$  are
interpreted  as the masses and the decay  widths of the  physical
states  $\tilde{K}_{1,2}$. These states are usually denoted as
$K_{L,S}$, but since we shall be representing the physical states 
including VLI effects with the same notation we shall refer to them 
as $\tilde{K}_{1,2}$. The transformation matrix $U_W$ which relates 
the states $\tilde{K}_{1,2}$ to the states $K^0,\bar{K}^0$ can be 
written as
\begin{equation}\label{UW}
U_W = \frac{1}{\sqrt{2(1 + |\tilde{\varepsilon} |^2)}}
e^{i \chi_W}
\pmatrix{(1 + \tilde{\varepsilon}) &  (1 + \tilde{\varepsilon}) \cr -
(1 - \tilde{\varepsilon}) 
&(1 - \tilde{\varepsilon})}\pmatrix{ e^{-i\beta_W} & 0 \cr 0 & e^{i\beta_W}} .
\end{equation}
We have assumed that there is no $CPT$ violation in the 
non-VLI part of the Hamiltonian, but only that $CP$ is violated,
parametrized by $\tilde{\varepsilon}$. The phases $\chi_W$ and $\beta_W$
can be eliminated by a redefinition of the $\tilde{K}_{1,2}$ states in such a
way that we have the usual formula:
\begin{equation}\label{Ktilde}
\tilde{K}_{\stackrel{1}{2}} = \frac{1}{\sqrt{2(1 + |\tilde{\varepsilon} |^2)}} 
\left[ 
(1 + \tilde{\varepsilon}) K^0 \mp (1 - \tilde{\varepsilon}) \bar{K}^0
\right] 
\end{equation}
For the VLI part if we assume that the velocity eigenstates
are orthogonal they are related to the $K^0,\bar{K}^0$ by a
unitary matrix $U_v$ which can be written in the general form
\begin{equation}\label{Uv}
U_v = e^{i\chi_v}
\pmatrix{ e^{-i\alpha_v} & 0 \cr 0 & e^{i\alpha_v}} 
\pmatrix{ \cos\theta_v & \sin\theta_v \cr -\sin\theta_v & \cos\theta_v} 
\pmatrix{ e^{-i\beta_v} & 0 \cr 0 & e^{i\beta_v}} .
\end{equation}
The phases $\chi_v$ and $\beta_v$ can be absorbed in a redefinition of
the velocity eigenstates. The phase $\alpha_v$ (which is 
similar to
{\rm Im}$\tilde{\varepsilon}$ in Eq.(\ref{UW}) which 
to this order in $\tilde{\varepsilon}$ can be written
in the form of such a phase) cannot be absorbed because $K^0$-$\bar{K}^0$
are by definition charge conjuguate states.
The phase $\alpha_v$ is a new source of $CP$ violation which can be present
even though the velocity states are still orthogonal. 

{}From the form of the transformation matrix $U_W$ and $U_v$, the total
hamiltonian in the $K^0-\bar{K}^0$ basis is
$$
H = p I + \frac{1}{2 p} { \pmatrix{ M_{+} & M_{12} \cr 
M_{21} & M_{-} }}^2
$$
with
\begin{eqnarray}
M_{\pm} &=& \bar{m} - i {\bar{\Gamma} \over 2}
\pm \frac{p^2}{\bar{m}} \frac{\cos 2\theta_v}{2}
\delta{v}  \nonumber \\
M_{12} &=& - {1 \over 2}\frac{1 + \tilde{\varepsilon}}
{1 - \tilde{\varepsilon}}
(\delta m - i {\delta \Gamma \over 2}) 
- e^{-2i \alpha_v}\frac{p^2}{\bar{m}} \frac{\sin 2\theta_v}{2}
\delta v \nonumber \\
M_{21} &=& - {1 \over 2}\frac{1 - \tilde{\varepsilon}}
{1 + \tilde{\varepsilon}}
(\delta m - i {\delta \Gamma \over 2}) 
- e^{2i \alpha_v}\frac{p^2}{\bar{m}} \frac{\sin 2\theta_v}{2}
\delta v \label{mtotv}
\end{eqnarray} 
The mass matrix above is the general formula from which we will
discuss different special cases.

\section{General mass matrix for VEP.}

To formulate the VEP mechanism in the Kaon system, 
we first study the energy of the  particles  under  consideration,
taking the kaons to be relativistic.  The 
gravitational part of the Lagrangian to first order (linearized 
theory) in a
weak gravitational field $g_{\mu\nu}=\eta_{\mu\nu}+    h_{\mu\nu}$   
(where   $h_{\mu\nu}= 2\frac{\phi}{c^2}  {\mbox diag}(1,1,1,1)$) 
can be written as ${\cal L} = -\frac{1}{2}(1+g_i)h_{\mu\nu}T^{\mu\nu}$
where  $T^{\mu\nu}$  is the  stress-energy  in the  gravitational
eigenbasis. The principle of equivalence says that the 
gravitational couplings $g_i$ are equal. 

We can now write down the effective Hamiltonian including the 
strong, electromagnetic, weak, and  gravitational interactions in the $(K^0
\ ,\bar{K}^0)$ basis :
\begin{equation}
H = p  I + U_W H_{SEW} U_W^{-1} + U_G H_G U_G^{-1} \label{hha}
\end{equation}
with I the identity matrix, 
\begin{equation}
H_{SEW} = \frac{(M_{SEW})^2}{2 p} = \frac{1}{2 p} {\pmatrix{ 
m_1 - i {\Gamma_1 \over 2} & 0 \cr 0 & m_2 - i {\Gamma_2 \over 2}}}^2 \label{hsewa}
\end{equation}
and
\begin{equation}
H_G = \pmatrix{ G_1 & 0 \cr 0 & G_2} = \pmatrix{ 
- 2 (1 + g_1) \phi (p + \frac{\bar{m}^2}{2 p}) & 0 \cr 0 &
- 2 (1 + g_2) \phi (p + \frac{\bar{m}^2}{2 p})} \label{hga}
\end{equation}
in physical time and length units \cite{hughes}
to first order in $\bar{m}^2/p^2$ with $p$ the momentum.
In formalisms where the weak equivalence principle is assumed 
\cite{Kenyon,hughes}, one 
starts with $U_G$ proportional to $U_W$ (in the case considered where
$CP$-violating effects in $U_W$ are taken to be 0), which leads to a 
violation of
$CPT$ if VEP is operative, that is to say if $g_1 \neq g_2$. More generally
when VEP is operative, $U_G$ is not necessarily proportional to $U_W$.
Note that in the gravitational Hamiltonian
$H_G$  we  have  neglected  terms   proportional  to
$\delta{m}$, and $\phi$ is the  gravitational  potential on the surface of
earth,   which  is  constant   over  the  range  of   terrestrial
experiments.

In the  absence of  gravity,  $m_{1,2}$  and  $\Gamma_{1,2}$  are
interpreted  as the masses and the decay  widths of the  physical
states  $\tilde{K}_{1,2}$ defined by Eq.(\ref{Ktilde}) as 
in section 2.
For the gravitational part if we assume that the gravitational states
are orthogonal they are related to the $K^0,\bar{K}^0$ by a
unitary matrix $U_G$ which can be written in the general form
\begin{equation}\label{UG}
U_G = e^{i\chi_G}
\pmatrix{ e^{-i\alpha_G} & 0 \cr 0 & e^{i\alpha_G}} 
\pmatrix{ \cos\theta_G & \sin\theta_G \cr -\sin\theta_G & \cos\theta_G} 
\pmatrix{ e^{-i\beta_G} & 0 \cr 0 & e^{i\beta_G}} .
\end{equation}
The phases $\chi_G$ and $\beta_G$ can be absorbed in a redefinition of
the gravitational states but the phase $\alpha_G$ cannot
like in the VLI case
\footnote{This is in contrast to the VEP mechanism for 
neutrinos \cite{utrobb,gasper,pantal} where
charge conjugation plays no role. Relative phases $\alpha$ between
neutrino flavour eigenstates can be absorbed in at least one sector, 
e.g. the weak sector \cite{pantal}.}
and is a new source of $CP$ violation like $\alpha_v$.

{}From $U_W$ and $U_G$ we then get 
\begin{eqnarray}
M_{\pm} &=& \bar{m} - i {\bar{\Gamma} \over 2}
+ \frac{p}{\bar{m}} \bar{G} \pm \frac{p}{\bar{m}} \frac{\cos 2\theta_G}{2}
\delta G  \nonumber \\
M_{12} &=& - {1 \over 2}\frac{1 + \tilde{\varepsilon}}
{1 - \tilde{\varepsilon}}
(\delta m - i {\delta \Gamma \over 2}) 
- e^{-2i \alpha_G}\frac{p}{\bar{m}} \frac{\sin 2\theta_G}{2}
\delta G \nonumber \\
M_{21} &=& - {1 \over 2}\frac{1 - \tilde{\varepsilon}}
{1 + \tilde{\varepsilon}}
(\delta m - i {\delta \Gamma \over 2}) 
- e^{2i \alpha_G}\frac{p}{\bar{m}} \frac{\sin 2\theta_G}{2}
\delta G \label{mtot}
\end{eqnarray} 
The mass matrix above is the general formula from which we will
discuss different cases like Eq.(\ref{mtotv}) in the VLI case. 

Comparing Eq.(\ref{mtotv}) and Eq.(\ref{mtot}) we see that both the
VLI and VEP mass matrices are similar. 
Puting $\theta_v$ = $\theta_G$ and $\alpha_v$ = $\alpha_G$, 
VLI and VEP effects are 
indistinguishable to lowest order in $\bar{m}^2/p^2$ 
providing one identifies the VLI parameter $\delta v$ with 
the VEP paramter $-2 \phi \delta g$. From
now on we will discuss the VEP case knowing that any
corresponding VLI formula can be obtained
straightforwatdly from this identification.

\section{Testing the equivalence principle in the case $\alpha = 0$.}

In this section we restrict ourselves to  tests of the equivalence
principle from $m_L -m_S$ data in the case where gravitational
states are related to the states $K^0-\bar{K}^0$ by a simple orthogonal
matrix (i.e. $\alpha_G$= 0).
In a first step we neglect the decay widths in Eq.(\ref{hsewa})-(\ref{mtot}).
In the basis of the 
physical states $K_L$ and $K_S$, the 
hamiltonian of Eq.(\ref{mtot}) becomes
\begin{equation}
H = \pmatrix{p + \frac{m_L^2}{2 p} & 0 \cr 0 & p + \frac{m_S^2}{2 p} }
\label{he}
\end{equation}
%
with $(m_L +m_S)/2=\bar{m} + \frac{p}{m} \bar{G}$
and
\begin{equation}
m_L - m_S = \left[ 
( \delta m )^2 + {\left( 2 \phi \delta g {p \over \bar{m}}
(p + {\bar{m}^2 \over 2 p}) \right)}^2 - 4 \delta m \phi \delta g
\frac{p}{\bar{m}} (p + {\bar{m}^2 \over 2 p} ) 
\cos(\frac{\pi}{2} - 2 \theta_G) \right]^{1 / 2} 
\label{mls}
\end{equation}
where $m_L$ and $m_S$ are the experimentally measured masses of 
$K_L$ and $K_S$ respectively. From this expression it is clear 
that the mass difference $m_L - 
m_S$ is energy dependent. (The possibility of energy dependence 
of the various parameters in the Kaon system has been 
previously considered in different contexts 
[3,9-13]).

{}From Eq.(\ref{mtot}) we can define the amount of $CPT$ violation induced by
VEP as follows
\begin{equation}
\Delta_{CPT} = M_{+} - M_{-} = - 
\cos (2  \theta_G) 2 \phi \delta g \frac{p}
{\bar{m}} (p + {\bar{m}^2 \over 2 p}) \label{dcpt} 
\end{equation}
Recent studies of VEP in the Kaon system \cite{Kenyon}-\cite{hughes}
assumed $CPT$ violation
in the gravitational sector, from which  
 it was argued that empirical bounds 
can be placed on the difference between the gravitational couplings
to $|K^0>$ and $|\bar{K}^0>$. 
The difference in gravitational 
eigenvalues then corresponds to a difference ($\Delta M_g$) 
in the masses of  $|K^0>$ and $|\bar{K}^0>$:
\begin{equation}
\vert M_{+} - M_{-}\vert = \phi \Delta M_g = 2
\phi \vert\delta g\vert \frac{p}{\bar{m}} (p + {\bar{m}^2 \over 2 p}) \label{kh}
\end{equation}
and is entirely attributable to the
amount of $CPT$ violation. The first equality in Eq.(\ref{kh}) 
was given by  Kenyon \cite{Kenyon} and the second by
Hughes \cite{hughes}, who specified the energy dependence 
of $\Delta M_g$. From the experimental upper bound on $M_{+} - M_{-}$ 
\cite{Carosi} the bound $\mid\delta g\mid  < 2.5\times10^{-18}$ may
be obtained, where the potential $\phi$ is taken to be that due to
the local supercluster ($\phi \simeq 3\times 10^{-5}$) and $p \simeq$
100 $GeV$ \cite{Carosi}.
In this approach CPT conservation implies
no gravitational mass difference and hence no VEP.
However it is clear from the expression (\ref{dcpt})
for $\Delta_{CPT}$ that the bound obtained 
on $\Delta M_g$ is actually on some combination of 
VEP parameters and
not on $\delta g$ and $\cos (2 \theta_G)$ separately.
When $\theta_G = 0$, Eq.(\ref{kh}) 
agrees with Eq.(\ref{dcpt}). More recent experiments \cite{pdg} find
$|M_{+}-M_{-}|/m_K < 9\times 10^{-19}$, yielding the bound
$\mid\delta g\mid < 3.8\times 10^{-19}$ for the same 
values of $\phi$ and $p$. 

In the case of VLI with $\theta_v$ = 0, the 
amount of $CPT$ violation associated with 
VLI is given by,
\begin{equation}
|M_+ - M_-| = |\delta v| {p^2 \over \bar{m}}
\end{equation}
The same experimental results can be used to constrain the VLI 
parameter: $|\delta v| < 2.3 \times 10^{-23}$. 
To leading order this has exactly the 
same energy dependence as the VEP mechanism.

Next we shall consider a scenario in which CPT is conserved,
so that $\Delta_{CPT} = 0$. From the above it is clear that, even if CPT
is conserved,
there is still a VEP-induced difference between the masses of 
the physical states.  As a result bounds can be placed
on the VEP parameter $\phi\delta g$ without the assumption that
locality in quantum field theory is violated. 

{}From the expression of $\Delta_{CPT} $ it is clear that it is
possible to conserve CPT for all momentum taking 
$\theta_G = {\pi \over 4}$ (modulo $\pi \over 2$).
In this case the mass difference is 
\begin{equation}
m_L - m_S =  \delta m - 2 \phi \delta g {{p} \over
 {\bar{m}}} (p + {\bar{m}^2 \over 2 p}) 
\label{bef}
\end{equation} 
which as noted above is energy dependent and to the leading 
order similar to the VLI expression in the case $\theta_v$ = $\pi /4$:
 $m_L - m_S$ = $\delta m$ + $\delta v p^2 /m$. 
It is possible to put a bound on the VEP parameter
$\delta g$ if we know the value of $\phi$ and 
the mass  difference at various given energies. Alternatively, if mass 
measurements at two different energies were different, the
differing values for $m_L - m_S$ could be used to
extract a value for the VEP parameter $\delta g$. 

We now proceed to find out constraints on the parameters 
$\delta m$ and $\delta g$ (or $\delta v$). In the review of 
particle properties \cite{pdg} six experiments were taken into 
account. Two of them \cite{gib,sch} are with the kaon momentum 
$p_K$ between 20 GeV and 160 GeV. The weighted average of these 
two experiments is \cite{sch}: $\Delta m_{LS} = m_L - m_S = 
(0.5282 \pm 0.0030) 10^{10} \hbar s^{-1}$. The four other 
experiments \cite{cullen,gew,gjes,adler} are at lower energy, 
with $p_K \approx 5$ GeV, or less with a  weighted average 
$\Delta m_{LS} = (0.5322 \pm 0.0018) 10^{10} \hbar s^{-1}$. A 
fit of equation (\ref{bef}) with the high and low energy value 
of $\Delta m_{LS}$ gives : $\delta m = (3.503 \pm 0.012) \times 
10^{-12} MeV$ and $\phi \delta g = (8.0 \pm 7.0) \times 
10^{-22} \times \left( \frac{90}{E_{av}} \right)^2$, (where 
$E_{av}$ is the average energy for the high energy experiment). 
All these bounds on the VEP parameter $\phi \delta g$ are also bounds
of the VLI parameter $-\delta v /2$ with the same 
energy dependence. We shall not explicitly present the VLI 
bounds. 

Taking $\phi$ to be the earth's potential ($\phi \simeq 
0.69\times 10^{-9}$), we find $\delta g = (1.2 \pm 1.0)\times 
10^{-12}$ whereas if $\phi$ is due to the local supercluster 
then $\delta g = (2.7 \pm 2.3)\times 10^{-17}$. These values 
differ from zero by 1.15 standard deviations. 
A precise 
fit of mass difference per energy bin in present and future 
high energy experiments would be extremely useful in 
constraining the energy dependent VEP or VLI parameters. 
Improvement on the low energy experiments can also change the 
bounds. One of the low energy experiments 
published last year found $\Delta m_{LS} = (0.5274 \pm 0.0029 
\pm 0.0005)\times 10^{10} \hbar s^{-1}$ \cite{adler}; when fitted with the 
high energy experiments, a value of $\delta g$ 
consistent with 0 at less than 1 standard deviation is obtained.
On the  other hand, without this new experiment, a similar fit of the 
other five experiments yields $\phi \delta g = (1.38 \pm 0.77) 
\times 10^{-21} (90/E_{av})^2$. In this case $\delta g$ is 
different from 0 by 1.8 standard deviations. 

In the above analysis we have not included the effect of the absorptive
part of the Hamiltonian, i.e of the decay widths in Eqs.(\ref{hsewa}-\ref{mtot}).
Including them we now obtain 
\begin{eqnarray}
m_L - m_S &=& {1 \over
 \sqrt{2}} \left[ \sqrt{F^2 + G^2} + F \right]^{1/2} 
\label{ls} \\
\Gamma_L -\Gamma_S &=& \sqrt{2} \left[ \sqrt{F^2 + G^2} - F \right]^{1/2} 
\label{ps} 
\end{eqnarray}
\begin{eqnarray}
F &=& (\delta m)^2 + (2 \phi \delta g {{p} \over {\bar{m}}}(p + 
{\bar{m}^2 \over 2 p} 
))^2 - 4 
\delta m \phi \delta g {{p} \over
 {\bar{m}}} (p + {\bar{m}^2 \over 2 p}) \cos (\frac{\pi}{2} - 2 \theta_G)
- ( {{\delta \Gamma} \over 2})^2 \nonumber \\
G &=& - ( \delta m  \delta \Gamma) + 2 \cos (\frac{\pi}{2} -
2 \theta_G) [ \delta \Gamma \phi \delta g {{p} \over {\bar{m}}} (p +
 {\bar{m}^2 \over 2p}) ] \nonumber
\end{eqnarray}
In deriving these equations we neglected terms in $\delta m  \Gamma$,
$\delta m \delta \Gamma$ and $\Gamma^2$ with respect to the terms in
$m \delta m$ or $m \delta \Gamma$. 
It can be shown that in the $CPT$-conserving case the mass
difference given in Eq. (\ref{ls}) reduces to Eq. (\ref{bef}).
So in the $CPT-$conserving case the results above are not affected
by inclusion of the widths. In this case the difference
$\Gamma_S - \Gamma_L = \delta \Gamma$ is independent of energy.
This is consistent with experiment, which indicates that
the low and high energy measurements of $\Gamma_S - \Gamma_L$ are
fully compatible \cite{pdg}. 
For $\theta_G \not= \pi/4$, an examination of (\ref{ps})
indicates that $\Gamma_S - \Gamma_L$ is energy dependent; however 
this is small and measurements of $\Gamma_S - \Gamma_L$ do not
constrain $\delta g$ more than measurements of $\Delta m_{LS}$ even
though they are relatively more precise. We note that
measurements of $\Gamma_S - \Gamma_L$ would more strongly constrain
a possible absorptive part coming from the gravitational sector
which presumably would induce a larger energy dependence.
We shall not consider this possibility here.
For $\theta_G \not= \pi/4$, width effects
in Eq.(\ref{ls}) are small and (\ref{mls}) remains valid to within
a few percent.

In Fig.1, for completeness, we plot as a function of $\cos (2\theta_G)$ 
the upper bounds we get 
on $\vert \phi \delta g\vert$ by fixing
$\delta m$ to the central value of the world average \cite{pdg},
$\Delta m_{LS}$ =$(0.5310 \pm 0.0019)\times 10^{10} \hbar s^{-1}$ and 
requiring that
$m_L -m_S$ in (\ref{mls}) does not differ from $\delta m$ by more that
$\pm 2$ standard deviations. Note that in the case of maximal 
$CPT$ violation ($\theta_G = 0$), $m_L-m_S$ can only increase with energy,
as is clear from Eq.(\ref{mls}) or Eq. (\ref{ls}). The actual difference
between low and high energy experiments, if valid, 
could not be explained in this case except for complex values of 
$\delta g$ (and similarly for values
of $\theta_G$ very close to 0). 

In Fig.1 we also show the bound coming
from Eq.(\ref{dcpt}) requiring that at high
experimental energy ($p \simeq 100 GeV$ ) the experimental 
upper bound $|M_{+}-M_{-}|/m_K < 9\times 10^{-19}$ \cite{pdg} 
on $CPT$ violation is satisfied.


\section{Testing the equivalence principle from $CP$ violation
experiments.}

Now we consider the effect of the $CP$-violating phase $\alpha_G$. Let
us note first that in the maximally $CPT-$violating case,
$\theta_G$ = 0, there is no effect of the phase
$\alpha_G$ in the mass matrix Eq.(\ref{mtot}) and consequently no 
$CP-$violating effect coming from this phase (similarly in the VLI case
when matter and antimatter states are the velocity eigenstates).
In the following we will restrict ourself to the most interesting 
$CPT-$conserving case with
$\theta_G = \pi/4$.
The total Hamiltonian can then be diagonalised 
\begin{equation}
H =  \pmatrix{ p+ {1 \over 2 p}(m_L - i {\Gamma_L \over 2})^2 & 0
\cr 0 & p+ {1 \over 2 p}(m_S - i {\Gamma_S \over 2})^2 }
\end{equation}
with the physical eigenstates $K_L$ and $K_S$ being given by
\begin{equation}
K_{\stackrel{L}{S}} = \frac{1}{\sqrt{2(1 + |{\varepsilon} |^2)}} \left[ 
(1 + {\varepsilon}) K^0 \mp (1 - {\varepsilon}) 
\bar{K}^0 \right] 
\end{equation}
Defining
\begin{eqnarray}
G_c &=& {p \over \bar{m}} \delta G \cos 2 \alpha_G =
- 2 g_c (\frac{p^2}{\bar{m}} + {\bar{m} \over 2} ); \quad 
\ g_c = \phi\delta g \cos 2 \alpha_G 
\nonumber \\
G_s &=& {p \over \bar{m}} \delta G \sin 2 \alpha_G =
- 2 g_s (\frac{p^2}{\bar{m}} + {\bar{m} \over 2} ); \quad
\ g_s = \phi\delta g \sin 2 \alpha_G 
\nonumber 
\end{eqnarray}
the new $CP$-violating parameter ${\varepsilon}$ is now defined 
in terms of the $CP-$violating parameters $\tilde{\varepsilon}$ and $G_s$ via
\begin{equation}
\label{eps}
{\varepsilon} =   
\frac{\tilde{\varepsilon} (\delta m  
- i {\delta \Gamma \over 2}) 
- {i \over 2} G_s 
}
{(\delta m + G_c)
- i {\delta \Gamma \over 2}} 
\end{equation}
to first order in $\tilde{\varepsilon}$ and $G_s$.
Similarly we have to first order in
$\tilde{\varepsilon}$ and $G_s$
\begin{eqnarray}
\label{Delm}
\Delta m \equiv m_L - m_S
&=& ( \delta m + G_c)
 \\
\label{DelG}
\Delta \Gamma \equiv \Gamma_L - \Gamma_S &=&  \delta \Gamma.
\end{eqnarray}

Since in the mechanism considered here there is no $\varepsilon^\prime$
type $CP$ violation coming from VEP  we will neglect
other possible $\varepsilon'$ effects and the relevant $CP$-violating
quantities are
$$\begin{array}{rcccl}
\delta &=& \displaystyle \frac{\Gamma(K_L \to \pi^- l^+ \nu) - 
\Gamma(K_L \to \pi^+ l^- \nu)}
{\Gamma(K_L \to \pi^- l^+ \nu) + \Gamma(K_L \to \pi^+ l^- \nu)}
&=& \displaystyle 2{\rm Re} \varepsilon  \\[.3in]
\eta_{+-}&=& \displaystyle \frac{A(K_L \to \pi^+ \pi^-)}{A(K_S \to \pi^+ \pi^-)} =
|\eta_{+-}| e^{i \phi_{+-}} &=& \varepsilon \\[.3in]
\eta_{00}&=& \displaystyle \frac{A(K_L \to \pi^0 \pi^0)}{A(K_S \to \pi^0 \pi^0)} =
|\eta_{00}| e^{i \phi_{00}}&=& \varepsilon \\[.3in]
\end{array}$$.

Consider first the case $\tilde{\varepsilon}=0$, {\it i.e.}
there is no $CP$ violation induced from the weak interaction.
Can we interpret the observed $CP$ violation parameters above
as originating purely due to the relative phase $\alpha_G$? 
In other words does the superweak mechanism have a gravitational
origin?
Eq. (\ref{eps}), with $\tilde{\varepsilon}$ = 0, can be written as
\begin{equation}
\varepsilon = - {i \over 2} {G_s \over 
( \Delta m - i{ \Delta \Gamma \over 2})} .
\end{equation}
Equating the real and the imaginary parts we get
$$ {\rm Re} \varepsilon = {G_s \over 2} {\Delta \Gamma \over 2} 
[(\Delta m)^2 + ({\Delta \Gamma \over 2})^2]^{-1}
\hskip .2in ; \hskip .3in 
{\rm Im} \varepsilon = -{G_s \over 2} {\Delta m} 
[(\Delta m)^2 + ({\Delta \Gamma \over 2})^2]^{-1} .$$
The above equations reproduce the results of the
superweak theory: $\phi_{+-}=\phi_{00} = 
-2\Delta m/\Delta \Gamma \simeq 43.5^0 $ and
also consequently 
$ |\eta_{+-}|
= |\eta_{00}| \simeq {\delta \over \sqrt{2}} $. 
The fact that the superweak phase is obtained is due to the fact that
the VEP mechanism considered here respects the hermiticity of the
interaction between the $CP-$eigenstates $K_1$ and $K_2$ (i.e. the
numerator of Eq.(\ref{eps}) is purely imaginary
\footnote{Note that insofar as there is no gravitational coupling 
to the imaginary part of the energy (as we have assumed here),
a $CP$ violation of the
{\rm Re}$\varepsilon$ type coming from gravitation (which would render
$U_G$ nonunitary) does not respect the hermiticity of the $K_1$-$K_2$
interaction  and is consequently experimentally
suppressed. We do not consider the possibility of such an effect here.}).
Interestingly, we see that by assuming the gravitational, weak and 
mass  eigenstates are all related by unitary transformations, 
the physical states are still of the superweak type; 
in particular they are no longer related to the other states by a 
unitary transformation and are no longer orthogonal.

 Taking the experimental value
$\delta = (0.327 \pm 0.012)\%$ \cite{pdg} as input we obtain
$G_s \simeq - 2 \times 10^{-14}$. This value of $G_s$ yields a consistent
fit to all the $CP$ violation parameters above, as with any
superweak mechanism.
However this does not provide us with positive evidence for 
VEP-induced $CP$ violation because
$|\eta_{+-}|$ and $|\eta_{00}|$ have been
observed experimentally with good accuracy to be constant over a 
large energy range \cite{pdg}. Hence 
is not possible to reproduce the data for all energies with
$\tilde{\varepsilon} = 0$ since $G_s$ is proportional to $p^2$.  
An observed energy dependence in these parameters that is
consistent with (\ref{eps}) would be a definitive signature of
a VEP-mechanism operative in this sector.

We now demonstrate how a bound on the VEP parameter $\phi \delta g$ can be 
obtained independently of the phase $\alpha_G$. 
We can extract
a bound on $g_c$ from the experimental constraint on the energy
dependence of $\Delta m$, Eq. (\ref{Delm}) in the same way that for $\phi
\delta g$ above in Eq.(\ref{bef}) substituting $\delta g$ by $\delta g \cos 2
\alpha_G$. The bound obtained on $\phi \delta g$ in Fig.1 is now a bound
on $g_c$:
\begin{equation}
|g_c| < 9 \times 10^{-22}.
\end{equation}
{}From $CP$ violation parameter data we can also obtain a bound on $g_s$.
Defining 
$$ A = {G_c \over {\delta m}} \hskip .3in B = {{\rm Re} \varepsilon \over
{\rm Im} \varepsilon} - 1 \hskip .3in C = - {\delta \Gamma \over 2 \delta m}
- 1 $$
where the magnitudes of $A$, $B$ and $C$ are all much smaller than unity,
implies from  (\ref{eps})  
\begin{eqnarray}
{\rm Re} \varepsilon &=& {\rm Re} \tilde{\varepsilon} ( 1 - A) +
{1 \over 2} G_s \frac{{\delta \Gamma \over 2}}{(\delta m)^2 +
({\delta \Gamma \over 2})^2}  \\
\label{Imeps}
{\rm Im} \varepsilon &=& {\rm Im} \tilde{\varepsilon} - {1 \over 2} G_s
\frac{\delta m}{(\delta m)^2 +
({\delta \Gamma \over 2})^2}  
\end{eqnarray}
where only terms linear in $A$, $B$ and $C$ have been retained. 
We observe that $g_s$ changes the value of $|\varepsilon|$ but not the
phase of $\varepsilon$. In addition, {\rm Im}$\varepsilon$ depends on 
$g_s$ but not on $g_c$.  From the
experimental values of  $\phi_{+-} = (43.7 \pm 0.6)^{\circ}$ and 
$|\eta_{+-}| = (2.284 \pm 0.018)10^{-3}$ \cite{pdg} we 
obtain {\rm Im}$\varepsilon = (1.58 \pm  0.02).10^{-3}$.
Fixing {\rm Im}$\tilde{\varepsilon}$ to
the central value of {\rm Im}$\varepsilon$ and requiring that 
{\rm Im}$\varepsilon$ in Eq.(\ref{Imeps}) differs from the central
value by no more than 2 
standard deviations at high experimental 
energies ($p \simeq 70$ GeV\cite{pdg}) yields
$$
|g_s| < 3 \times 10^{-23}
$$
This value hardly varies when we calculate {\rm Im}$\varepsilon$ from
the experimental value of $|\eta_{00}|$ and $\phi_{00}$ 
instead of $|\eta_{+-}|$ 
and $\phi_{+-}$. From the bounds on $g_c$ and $g_s$ we 
then get 
$$
|\phi \delta g| < 9
\times 10^{-22}.
$$
{}
$g_c$ can also be bounded from $CP$ violation. Indeed the $CP$-conserving
parameter $g_c$ is present in {\rm Re}$\varepsilon$ and a bound on it
can be obtained by considering the phase of 
$\varepsilon$
\begin{equation}
\tan \phi_{+-}=\frac{{\rm Im}\varepsilon}{{\rm Re}\varepsilon}=
\frac{{\rm Im}\tilde{\varepsilon}}{{\rm Re}\tilde{\varepsilon}}(1+A)
\end{equation}
which doesn't depend on $g_s$.
Taking the low energy value of $\phi_{+-}$
to equal its central
value above we similarly obtain (requiring that $\phi_{+-}$ not
differ at high energy by more than two standard deviations from its low
energy value): 
$$
|g_c| < 7 \times 10^{-21}.
$$ 
This value is of the same
order of magnitude as the upper bound obtained above by looking for 
energy dependence in $\Delta m$. 
$CP$ violation measurements consequently are a useful means for
searching both for $CP-$conserving VEP effects (through the
parameter $g_c$) and $CP-$violating VEP effects (through $g_s$).

We shall not consider the case where $CPT$ is violated ($\theta_G \neq
\pi/4$) with $\alpha_G \neq 0$. 
Equations
similar to Eqs.(\ref{eps})-(\ref{DelG}) can be straightforwardly
obtained from Eq.(\ref{mtot}) but they are lengthy and do not
provide any new interesting
physical results which have not already been discussed above. 

In ref.\cite{nacht}, the effect of a tensorial 
field $f_{ij}^{\mu \nu}$ whose $CPT-$violating 
interactions with the kaons is given by the
lagrangian {\rm L}=$f_{ij}^{\mu \nu}d_{\mu}\phi_i d_{\nu} \phi_j$ has
been considered with $\phi_{1(2)}$ the $CP$-eigenstates and $f_{11}^{\mu
\nu}$, $f_{22}^{\mu \nu}$,
$f_{12}^{\mu \nu}=f_{21}^{\mu \nu}$ real parameters. Writing $f_{ii}^{\mu
\nu}$ as $\frac{f_i}{f}\phi \eta^{\mu \nu}$ and $f_{12}^{\mu \nu}$ as
$\frac{f_T}{f}\phi \eta^{\mu \nu}$, constraints on 
$(f_1-f_2)/f$ and $f_T/f$ have been obtained from experimental energy
constraints on the energy dependence of $\Delta m_{LS}$ and 
$\eta_{+-}$ respectively.
We observe that, except for a different experimental situation, 
the constraints obtained on $|(f_1 - f_2)/f|$ and $|f_t/f|$ 
are similar to ours on the
corresponding quantities, $|4\delta g \cos 2 \alpha_G|$ and $|\delta g \sin
2\alpha_G|$ respectively with $\theta_G$ = $\pi /4$. From our treatment, we see consequently that
provided the tensorial interaction is of gravitational origin the
bounds on the parameters $|(f_1 - f_2)/f|$ and $|f_t/f|$ are in fact
bounds on a combination of the VEP parameters, the difference $\delta g$ in the
gravitational couplings and the phase $\alpha_G$.

We close this section by noting that the bounds 
on $| g_c |$, $| g_s |$ and $| \phi \delta g |$ 
above are also bounds on
the corresponding VLI parameters $| \frac{\delta v}{2} \cos 2\alpha_v |$,
$| \frac{\delta v}{2} \sin 2\alpha_v |$ and $| \frac{\delta v}{2} |$
respectively.

\section{Summary and Conclusion.}

The Kaon system provides us with an interesting physical situation in
which we can empirically check the validity of special relativity and/or
the equivalence principle in a matter/antimatter sector of the standard 
model that includes 2nd generation matter.  A variety of interesting 
combinations of VLI/VEP effects exist which
can be associated with $CP$ violation and/or $CPT$ violation.
Violations of the equivalence principle in the Kaon system need not
violate CPT (which in turn implies a loss of locality in quantum 
field theory) as considered in recent studies. 

A general feature of the VLI/VEP mechanisms is that they
predict an energy dependence in $m_L - m_S$ and in the 
$CP$ violation parameters which can be empirically tested
to obtain bounds on the relevant parameters (such as $\phi \delta g$
for VEP or $\delta v$ for VLI). Since both VEP and VLI have the 
same energy dependence, although we can obtain bounds for both, 
it will not be possible to experimentally distinguish between 
the two mechanisms. 
Under the assumption that all such parameters are within
two standard deviations over the energy scales at which they
have been measured, present experiments provide rather stringent
bounds on the VEP (or VLI) mechanism. A more systematic search for 
energy dependence in $m_L -m_S$ and in the $CP$-violating parameters (such as
$\varepsilon$) will provide us with more definitive information
about the VEP (or VLI) mechanism in this sector. In
addition in our formalism, we observe that the $CPT$-conserving case
\cite{nacht} and the
$CPT$-violating case of recent VEP studies \cite{Kenyon,hughes} are
special cases of the same general mechanism.

\vspace{1cm}

\noindent
{\bf Acknowledgements}

This work was supported in part by the Natural Sciences and
Engineering Research Council of Canada.

\vskip 1.5in

Fig.1: Upper bounds on $\vert \phi \delta g\vert$ (for p $\simeq 90 GeV$)
obtained from Eq.(\ref{mls}) (solid line), Eq.(\ref{dcpt})(dashed line)
as explained in the text.

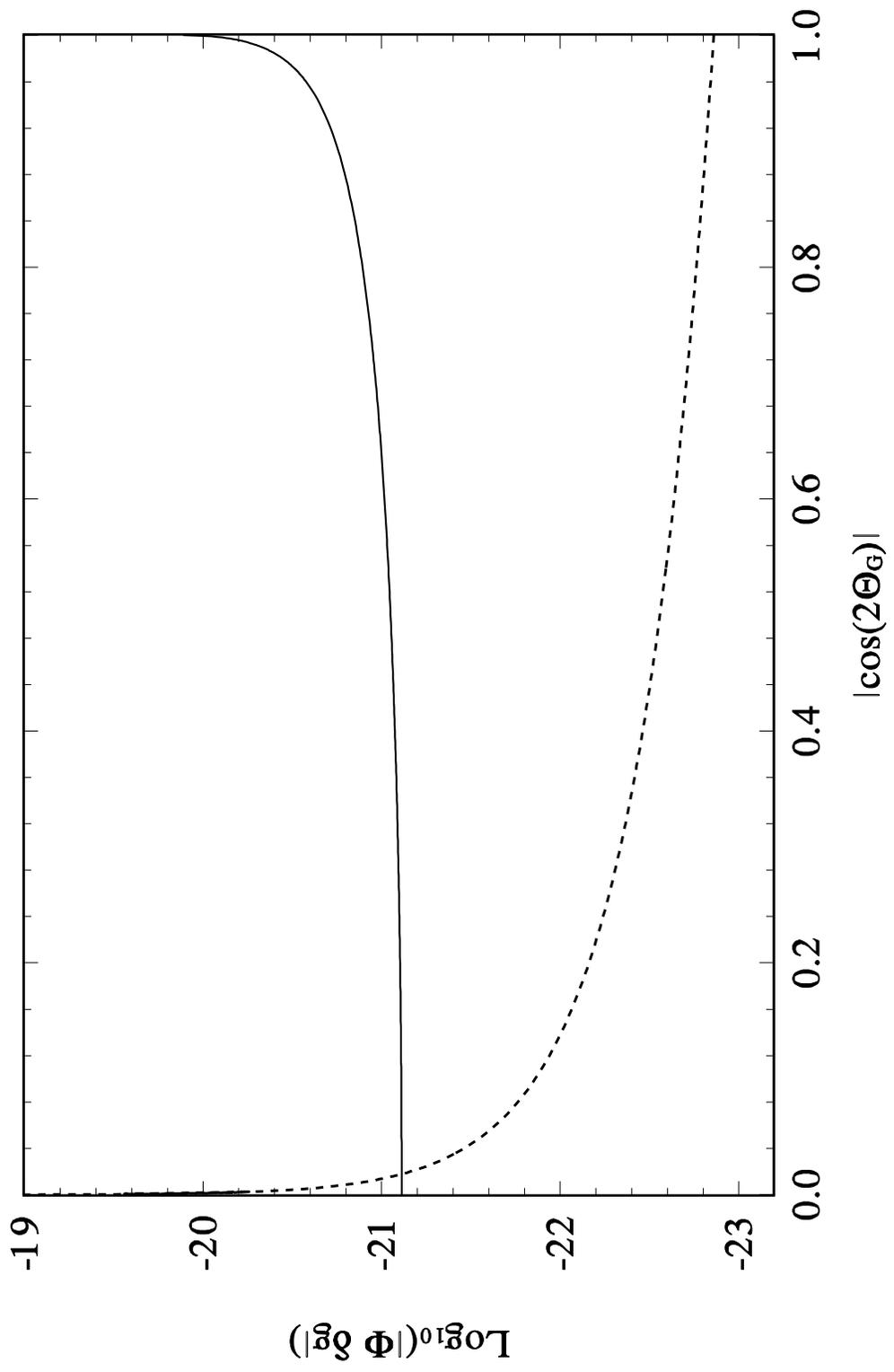
\end{document}